\documentclass[12pt]{article}

\usepackage{amsfonts}
\usepackage{amssymb}
\usepackage{amsthm}
\usepackage[fleqn]{amsmath}
\usepackage[mathcal]{euscript}
\usepackage{mathrsfs}

\input{epsf}

\newtheorem{proposition}{Proposition}

\newcommand{\Ref}[1]{(\ref{#1})}

\newcommand{\sign}{\mathrm{sign}\,}

\newcommand{\veps}{\varepsilon}

\newcommand{\vect}[1] {\boldsymbol{{ #1}} }

\newcommand{\qV}{{\vect{q}}}           

\DeclareMathAlphabet{\mathpzc}{OT1}{pzc}{m}{it}

\newcommand{\oli}[1]{\overline #1 }

\newcommand{\Rset}{\mathbb{R}}
\newcommand{\Sset}{\mathbb{S}}
\newcommand{\Tset}{\mathbb{T}}

\newcommand{\cE}{{\cal E}}

\begin{document}

\title{A note on classical ground state energies}

\vspace{-0.3cm}
\author{\normalsize \sc{Michael K.-H. Kiessling}\\[-0.1cm]
	\normalsize Department of Mathematics, Rutgers University\\[-0.1cm]
	\normalsize Piscataway NJ 08854, USA}
\vspace{-0.3cm}
\date{$\phantom{nix}$}
\maketitle
\vspace{-1.6cm}

\begin{abstract}
\noindent
	The pair-specific ground state energy $\veps_g(N):= \cE_g(N)/(N(N-1))$ of Newtonian $N$ body systems
grows monotonically in $N$.
	This furnishes a whole family of simple new tests
for minimality of putative ground state energies $\cE_g^{x}(N)$ obtained through computer experiments. 
	Inspection of several publicly available lists of such computer-experimentally obtained 
putative ground state energies $\cE_g^{x}(N)$  has yielded several dozen instances 
of $\cE_g^x(N)$ which failed one of these tests; i.e., for those $N$ one concludes that 
$\cE_g^x(N)>\cE_g(N)$ strictly. 
	Although the correct $\cE_g(N)$ is not revealed by this method, it does yield a better upper
bound on $\cE_g(N)$ than $\cE_g^x(N)$ whenever $\cE_g^x(N)$ fails a monotonicity test.
	The surveyed $N$-body systems include in particular $N$ point charges with $2$- or $3$-dimensional 
Coulomb pair interactions, placed either on the unit 2-sphere or on a 2-torus 
(a.k.a. Thomson, Fekete, or Riesz problems). 
\end{abstract}

\vfill
\hrule
\smallskip\noindent
{\footnotesize
Typeset in \LaTeX\ by the author.  Version of May 28, 2009. To appear in J. Stat. Phys.

\smallskip\noindent
\copyright 2009 The author. This preprint may be reproduced for noncommercial purposes.}
\newpage

\section{Introduction}
\noindent
	The pair-specific ground state energy $\veps_g(N)$ of Newtonian $N$ body systems
exhibits the following monotonic dependence on $N$: 
\begin{proposition}
\label{prop:UperNsqMinusN}
	Let $\Lambda\subset\Rset^{\rm d}$ be a bounded and connected domain, and let $\qV_k\in\oli\Lambda$.
	Assume the following hypotheses on $U_\Lambda(\check{\qV},\hat{\qV})$:

\smallskip\noindent
\begin{eqnarray}
&&(H1)\quad   {\mbox{\textit{Symmetry}:}}
		\hskip2.5truecm U_\Lambda(\check{\qV},\hat{\qV})=U_\Lambda(\hat{\qV},\check{\qV}) 
\nonumber\\
&&(H2)\quad   {\mbox{\textit{Lower\ Semi-Continuity}:}}
	\ U_\Lambda(\check{\qV},\hat{\qV})\ {\rm is\ l.s.c.\ on\ }  \oli{\Lambda}\!\times\!\oli{\Lambda}. 
\nonumber
\end{eqnarray}
	For $N\geq 2$, define the pair-specific ground state energy by 
\begin{equation} 
\veps_g(N)
\equiv
{\textstyle{\frac{1}{N(N-1)}}}
\min_{\{\qV_1,\dots,\qV_N\}}
\sum\sum_{\hskip-.7truecm 1\leq i < j\leq N} {U}_\Lambda(\qV_i,\qV_j).
\label{pairSPECpotENERGY}
\end{equation}	
	Then the sequence $N\mapsto \veps_g(N)$ so defined is monotonic increasing.
\end{proposition}

	For the convenience of the reader we here reproduce the elementary proof from Appendix A in \cite{KieRMP}.

\smallskip\noindent
{\textit{Proof of Proposition \ref{prop:UperNsqMinusN}:}} 

	We begin by noting that under hypotheses $(H1)$ and $(H2)$ the
pair-specific ground state energy $\veps_g(N)$ defined in \Ref{pairSPECpotENERGY} is well-defined;
i.e. $\veps_g(N)\in\Rset$.

	To prove the monotonicity of $N\mapsto \veps_g(N)$, with $N\geq2$,
let $\cE_g(N)$ denote the $N$-body ground state energy, i.e.
$\veps_g(N) = \cE_g(N)/(N(N-1))$.
	Using the definition of $\cE_g(N)$ and the elementary graph-theoretical identity that the sum over
all bonds in a complete $N$-graph ($N>2$) equals $(N-2)^{-1}$ the sum over all bonds of all its complete
$N-1$-subgraphs, and using the single inequality that the minimum of a sum is
not less than the sum of the minima, we find
\begin{eqnarray} 
\cE_g(N+1)
\!\!\!\!\!&& =
\min_{\ \{\qV_1,\dots,\qV_{N+1}\}}
\sum\sum_{\hskip-.7truecm 1\leq i < j\leq N+1} {U}_\Lambda(\qV_i,\qV_j)
\nonumber\\
\!\!\!\!\!&&  =
 \min_{\ \{\qV_1,\dots,\qV_{N+1}\}}
{\textstyle{\frac{1}{N-1}}}\!\!\!\!\!\sum_{\quad 1\leq k \leq N+1}
\Biggl[
\sum\sum_{\hskip-.7truecm \genfrac{}{}{0pt}{}{1\leq i < j\leq N+1}{ i\neq k\neq j} } {U}_\Lambda(\qV_i,\qV_j)
\Biggr]
\nonumber\\
\!\!\!\!\!&& \geq
{\textstyle{\frac{1}{N-1}}}\!\!\!\!\!\sum_{\quad 1\leq k \leq N+1}
\Biggl[
\min_{\{\qV_1,\dots,\qV_{N+1}\}\backslash\{\qV_k\}}
\sum\sum_{\hskip-.7truecm \genfrac{}{}{0pt}{}{1\leq i < j\leq N+1}{ i\neq k\neq j} } {U}_\Lambda(\qV_i,\qV_j)
\Biggr]
\nonumber\\
\!\!\!\!\!&& =
{\textstyle{\frac{N+1}{N-1}}} 
\min_{\{\qV_1,\dots,\qV_{N}\}}
\sum\sum_{\hskip-.7truecm {1\leq i < j\leq N}} {U}_\Lambda(\qV_i,\qV_j)
\nonumber\\
\!\!\!\!\!&&=\
{\textstyle{\frac{N+1}{N-1}}}\,\cE_g(N).
\label{potENERGYmono}
\end{eqnarray}
	Dividing \Ref{potENERGYmono} by $(N+1)N$ yields  $\veps_g(N+1)\geq \veps_g(N)$, 
and the proof of the monotonicity of $N\mapsto\veps_g(N)$ is complete.
\qed

	As already remarked in Appendix A of \cite{KieRMP}, Proposition \ref{prop:UperNsqMinusN} and its proof
are quite elementary and presumably known, yet after a serious search in the pertinent literature I came up empty-handed, and additional consultation with several of my local expert colleagues have given me the
impression that Proposition \ref{prop:UperNsqMinusN} is perhaps not known, and in any event not widely known.

	In this brief note we will be concerned with a very practical application of Proposition 
\ref{prop:UperNsqMinusN} which certainly is not generally known, as we will demonstrate.
	Namely, the  monotonic increase with $N$ of the true pair-specific ground state energies $\veps_g(N)$ 
furnishes a whole family of \emph{necessary criteria for minimality} which any empirical list of computer-experimental data $\{\cE_g^x(N); N=2,3,4,...,N_*\}$ for such ground state energies needs to satisfy; 
put differently, we have the following \emph{sufficient criterion for failure to be minimal}: 
\begin{equation}
\forall N\geq 2:\, ( \exists\, n\geq 1\, :\ \veps_g^x(N+n) < \veps_g^x(N)\Longrightarrow \cE_g^x(N) > \cE_g(N)).
\label{FAILUREcrit}
\end{equation}
	For each $n\geq 1$ one can use \Ref{FAILUREcrit} as a test for any computer-experimentally produced
list of putative ground state energies $\{\cE_g^x(N); N= 2,3,4,...,N_*\}$.
	Of course, if any particular $\cE_g^x(N)$ passes the $n$-th test for each $n \leq N_*-N$,
i.e. if $\veps_g^x(N+n) \geq \veps_g^x(N)\,\forall\, n=1,...,N_*-N$, this does not mean
that this $\cE_g^x(N)$ is a true ground state energy; only \emph{failing a test is significant}, for
\Ref{FAILUREcrit} asserts that \emph{a test-failing $\cE_g^x(N)$ is surely not a ground state energy}.
	Moreover, whenever $\cE_g^x(N)$ fails any such test, then by Proposition \ref{prop:UperNsqMinusN}
we also get a better empirical upper bound on the true ground state energy $\cE_g(N)$ than $\cE_g^x(N)$
from the remaining data of the computer-generated list $\{\cE_g^x(N); N=2,3,...,N_*\}$:
\begin{equation}
 \cE_g(N) 
\leq 
\min_{\quad 1\leq n\leq N_*-N}\left\{ \textstyle{\frac{N(N-1)}{(N+n)(N+n-1)}}\, \cE_g^x(N+n) \right\}.
\label{UPPERboundEgN}
\end{equation}
	Note that \Ref{UPPERboundEgN} is always true, but it only leads to a better empirical upper bound 
on $\cE_g(N)$ than $\cE_g^x(N)$ when $\cE_g^x(N)$ fails a test. 
	Note also that the computer-experimental data supply empirical upper bounds to the actual ground state 
energies within the numerical accuracy,\footnote{It is assumed here that the algorithms have computed
	the energy of \emph{some} configuration.}
i.e. we always have $\cE_g^x(N)\geq \cE_g(N)$.

	Subjecting some publically available lists of computer-generated data of putative ground state energies 
$\cE_g^{x}(N)$ for various $N$-body systems to the above tests has yielded several dozen instances of $\cE_g^x(N)$ 
which failed one of these tests; i.e., for those $N$ we conclude that $\cE_g^x(N)>\cE_g(N)$ strictly, and we get an 
improved empirical upper bound on $\cE_g(N)$ through \Ref{UPPERboundEgN}.
	The surveyed $N$-body systems are $N$ point charges with ${\rm D}$-dimensional Coulomb 
pair interactions (${\rm D}=2;3$), placed either on the unit 2-sphere or on a 2-torus 
(variably known as (elliptic) Fekete, Thomson, and Riesz problems). 
	The analysis of the sphere data is reported in the next section; for the torus data see section 3.

\section{Many point charges on the 2-sphere}
\vskip-.3truecm
\noindent
	Finding the minimum energy configuration(s) of $N$ point charges placed on the unit 2 sphere
$\Sset^2$ is a beautiful, intriguingly rich, and hard mathematical problem which in addition is relevant
to many fields of science; see the survey articles \cite{ErberHockneyTWO}, \cite{SaffKuijlaars},
\cite{AtiyahSutcliffe}, and \cite{HardinSaffONE}, and the website \cite{Womersley}.
	One can either interpret $\Sset^2$ as a two-dimensional ``physical space'' in its own right with 
$U_{\Sset^2}(\qV_1,\qV_2)$ given by the ${\rm D}=2$-dimensional Coulomb pair interaction 
$- \ln  |\qV_1-\qV_2|$, where $|\qV_1-\qV_2|$ is the cordal distance on $\Sset^2$;
incidentally, the cordal distance on $\Sset^2$ coincides with the three-dimensional Euclidean 
distance between two points $\qV_1$ and $\qV_2$ on $\Sset^2\subset\Rset^3$, but the embedding can be avoided
in the discussion. 
	Or, one can interpret $\Sset^2\subset\Rset^{\rm D}$ for ${\rm D}>2$ as a proper submanifold, with 
$U_{\Sset^2}(\qV_1,\qV_2)$ given by the ${\rm D}$-dimensional Coulomb pair interaction
$|\qV_1-\qV_2|^{2-D}$, where $|\qV_1-\qV_2|$ is the ${\rm D}$-dimensional Euclidean 
distance between two points $\qV_1$ and $\qV_2$ on $\Sset^2\subset\Rset^{\rm D}$.
	For small $N$ the ground state configuration (a.k.a. an $N$-tuple of Fekete points) can easily be
characterized explicitly,\footnote{For ${\rm D}=2\& 3$ the first dozen minimizers 
		are discussed in the delightful article \cite{AtiyahSutcliffe}.}
and the asymptotic large $N$-dependence of $\cE_g(N)$ can be, and to some extent has been \cite{RSZa}
determined analytically without seeking the exact Fekete points,\footnote{It ``suffices'' 
		to know that for large $N$ the Voronoi cells around the charges are mostly
		hexagons of a certain size;
		see \cite{SaffKuijlaars} for an enlightening discussion.
		We also remark that the leading order term in the asymptotics can be determined with
		the help of a very general variational principle, see \cite{KieSpoCMP}.}
but in general the problem defies analytical treatment.
	Computer experiments (e.g. \cite{RSZa}, \cite{RSZb},
\cite{AWRTSDW}, \cite{PerezGetal}, \cite{ErberHockneyTWO}, \cite{BCNTlett}, \cite{BCNTfull}, \cite{BCEG})
help finding candidates for the minimizing configuration and in any event yield empirical upper bounds $\cE_g^x(N)$ 
on the ground state energy $\cE_g(N)$.
	But even computers are soon overwhelmed because the number of local minimum energy 
configurations which are not global seems to grow exponentially with $N$ \cite{ErberHockneyTWO} so that
a computer algorithm is more and more likely to find one of these non-global minima. 
	Indeed, our tests have successfully detected a couple dozen non-global minimum energy 
values because their pair-specific value surpassed an ensuing empirical pair-specific energy value in 
some computer-generated list. 

\subsection{Two-dimensional Coulomb interactions}
\noindent
	Tables of computer-experimental ground state energies $\{\cE_g^x(N); N=2,3,4,...\}$ for $N$ point charges 
on $\Sset^2$ with two-dimensional Coulomb pair interaction $U_{\Sset^2}(\qV_1,\qV_2) =- \ln  |\qV_1-\qV_2|$ 
can be found on the website \cite{BCM}, more precisely:

http://physics.syr.edu/condensedmatter/thomson/shells/...

...shelltable.php?topology=sphere\&potential=0\&start=0\&end=5000,

\noindent
and in \cite{RSZb}, which contains many references to earlier studies.

	A value of $\cE_g^x(N)$ is listed in \cite{RSZb} for 
199 consecutive values of $N$ starting at $N=2$; one also finds there three additional $\cE_g^x(N)$ 
values for $N$ which correspond to the icosahedral group.
	On p.116 of their article, the authors describe various careful tests which they have undertaken 
to increase the chances of their $\cE_g^x(N)$ actually being $\cE_g(N)$.
	In addition to their tests I now have subjected their list of data $\{\cE_g^x(N); N=2,3,...,200\}$ to  
the $n=1$ monotonicity test (which, apparently, is not among the tests which the authors of \cite{RSZb} have
used).
	Their data passed the $n=1$ test (there is no point, then, to test with $n>1$).
	
	While all tables in \cite{RSZb} are of course permanent, those at \cite{BCM}, 
according to these authors, are interactive and are updated whenever some user finds a new and better pair $(N,\cE_g^x(N))$. 
	Therefore it is mandatory to also give the information on which day one downloaded the data from 
\cite{BCM} for study, which for the computer-experimental data on the ${\rm D}=2$ Coulomb ground state energies
I did on Feb. 21, 2009; yet, on Feb. 24, 2009 the crucial data I will be talking about below were still the same.
	 At the time of writing of this paper, this data list $\{\cE_g^x(N); N\in\{3,...,5000\}\}$ has (had) 
plenty of gaps, i.e. putative ground state energies $\cE_g^x(N)$ are absent for many values of $N$.
	While the absence of the case $N=2$ is not a problem, as $\cE_g(2)$ is exactly known, the first real
gap is the absence of any $\cE_g^x(11)$, any $\cE_g^x(19)$, and soon the gaps become larger and larger.
	The larger the gaps, the less likely one is to detect non-globally minimizing $\cE_g^x(N)$ with the 
monotonicity tests, yet two data points ``got caught in this net'':
$\cE_g^x(97) = -891.653265231 > \cE_g(97)$, 
for 
$\veps_g^x(100)-\veps_g^x(97)=-0.013678811 <0$;
and
$\cE_g^x(2000)= -386,187.080630499 >\cE_g(2000)$, 
for 
$\veps_g^x(4212)-\veps_g^x(2000)=-0.000503199<0$.
	Notice that while there is only a gap of 2 missing data between $\cE_g^x(97)$ and $\cE_g^x(100)$,
the gap between $\cE_g^x(2000)$ and $\cE_g^x(4212)$ is a whopping 2211.
	Since the non-globally minimizing $\cE_g^x(2000)$ was detected by the $n=2212$ test, it must be
quite far away from the actual ground state energy $\cE_g(2000)$.
	By \Ref{UPPERboundEgN}, and using $\cE_g^x(4212)= -1,722,205.927290610$ from \cite{BCM}, we find
the  upper bound,
\begin{equation}
 \cE_g(2000) 
\leq 
\textstyle{\frac{2000\cdot 1999}{4212\cdot 4211}}\, \cE_g^x(4212) = -388,198.8687.
\label{UPPERboundEg2000}
\end{equation}
	Similarly, with $\cE_g^x(100)= -1,083.376338235$ from \cite{BCM}, \!\Ref{UPPERboundEgN} with $\!n\!=\!3\!$
yields $\cE_g(97) \leq \textstyle{\frac{97\cdot 96}{100\cdot 99}}\, \cE_g^x(100) = -1,019.030349$; 
however, this upper bound is 
certainly beaten by $\cE_g^x(97)= - 1,022.023977757$ in \cite{RSZb}. 
	Incidentally, several other non-globally minimizing data $\cE_g^x(N)$ in the list at \cite{BCM}
which actually pass the monotonicity tests can be detected by simply comparing with the list in \cite{RSZb}
(at the time of writing, the non-global data at \cite{BCM} are $\cE_g^x(12)$, $\cE_g^x(36)$, $\cE_g^x(60)$, 
$\cE_g^x(87)$, $\cE_g^x(96)$, $\cE_g^x(100)$); yet, no such comparison is possible for $\cE_g^x(2000)$ as the data 
reported in \cite{RSZb} do not go beyond $N=282$.

	We end this subsection with an illustration of the monotonicity of the sequence $N\mapsto\veps_g(N)$ by 
plotting the monotonically increasing sequence $N\mapsto\veps_g^x(N)\geq \veps_g(N)$ using the data in \cite{RSZb}, 
with ``$=$'' for $N\leq 15$ \cite{AtiyahSutcliffe}.
\bigskip

\epsfxsize=13.5cm
\epsfysize=4.5cm
\centerline{\epsffile{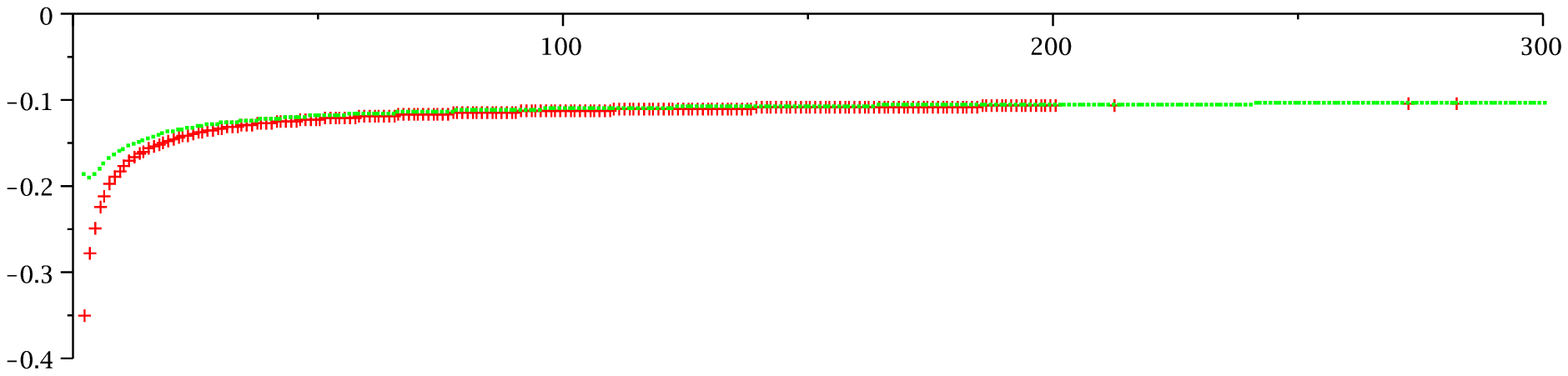}}
\begin{center}
{\scriptsize{
Computer-experimental pair-specific ground state energies $\veps_g^x(N)$ for $N$ point charges \\
on $\Sset^2$ with $\log r^{-1}$ Coulomb interactions, using the data of \cite{RSZb} (crosses). 
Shown \\
\vskip-.2truecm
also in this plot is the large $N$ asymptotic two-term approximation to $\veps_g(N)$ (dots).$\phantom{fi}$}}
\end{center}
\noindent
	Already for $N>50$ this empirical curve seems to agree to within less than $1\%$ absolute error with the 
large $N$ asymptotics of the actual sequence $N\mapsto \veps_g(N)$, given by
$\veps_g(N)\asymp 
	\textstyle{\frac{1}{4}}\ln \textstyle{\frac{e}{4}} - \textstyle{\frac{1}{4}} N^{-1}\ln N +O(N^{-1})$ 
and obtained through dividing by $N(N-1)$ the large-$N$ asymptotic formula for $N\mapsto\cE_g(N)$
\cite{RSZa,RSZb},
\begin{equation}
 \cE_g(N) = a N^2 + b N\ln N +O(N) 
\end{equation}
with $a= \textstyle{\frac{1}{4}}\ln\textstyle{\frac{e}{4}}$ and $b = - \textstyle{\frac{1}{4}}$.
	In \cite{RSZa} it is also conjectured that, actually, 
\begin{equation}
 \cE_g(N) = a N^2 + b N\ln N +cN +d\ln N + O(1),
\label{asympCONJlogS2}
\end{equation}
and rigorous upper and lower bounds on $c$ are given there.
	Smale's $7^{th}$ problem for the 21$^{st}$ century asks for an algorithm
which upon input $N$ returns an $N$-point configuration on $\Sset^2$ for which $\cE_g^x(N)$ does not deviate
from $\cE_g(N)$ by more than the fourth term in the conjectured expansion \Ref{asympCONJlogS2} (possibly up to
a different coefficient $d'$), and which does so in polynomial time; see \cite{Smale}.

\subsection{Three-dimensional Coulomb interactions}

	We next discuss the data for $N$ point charges  on the sphere $\Sset^2$ with  
three-dimensional Coulomb interactions $U_{\Sset^2}(\qV_1,\qV_2) = 1/|\qV_1-\qV_2|$, since \cite{Whyte}
referred to as ``Thomson's problem'' even though Thomson's ``plum pudding model of the atom'' \cite{Thomson} 
is not quite the same problem. 
	Be that as it may, there seem to be many more studies of this Thomson problem than of
its variants with other pair interactions and geometries.
	I perused a sample of those studies and eventually found data which failed a monotonicity test.

	Starting at $N=2$, in \cite{ErberHockneyTWO} one finds 111, and in \cite{RSZb} 199 consecutively 
computed putative ground state energies $\cE_g^x(N)$ for the Thomson problem (as just defined). 
	Since the authors in \cite{RSZb} mention that within numerical precision their data agree with 
those on Sloan's home page \cite{Sloanetal} (actually, at the time: its predecessor via netlib.att.com), 
I opted for analyzing the data on \cite{Sloanetal} which are stored (much) more userfriendly than those
in \cite{RSZb} and \cite{ErberHockneyTWO}, even though for only 129 consecutive values of $N$ some value 
$\cE_g^x(N)$ is listed, starting with $N=4$, plus a handful of other values for $N$s up to $N=282$. 
	All data at \cite{Sloanetal} passed the $n=1$ monotonicity test. 

	A wonderful treasure of data for the Thomson problem is listed on the website \cite{BCM},  
more precisely:

http://physics.syr.edu/condensedmatter/thomson/shells/...

...shelltable.php?topology=sphere\&potential=1\&start=0\&end=5000.

\noindent
	I downloaded the data on Feb.09, 2009, and on Feb.24 the relevant data were still present.
	Starting with $N=3$, one finds computer-experimental ground state energies $\cE_g^x(N)$ for an
amazing 1488 consecutive values of $N$, the first gap occurring at $N=1491$.
	In addition one finds almost 300 wider spaced $\cE_g^x(N)$ up to $N=5000$.
	Quite remarkably, all the consecutive values up to $N=1490$ passed the monotonicity test, and so
did an additional 70 values of $\cE_g^x(N)$ which are listed with gaps up to $N=1800$.
	The computer-experimental $\cE_g^x(1801)=1,579,605.0292504800$ is the first in the list which 
failed a monotonicity test, in fact the $n=1$ test, with $\veps_g^x(1802)-\veps_g^x(1801) = -0.0000044325$.
	Of course, there is also a better upper bound on $\cE_g(1801)$ via \Ref{UPPERboundEgN}, but 
it is not much of an improvement. 
	The story is quite different for the next failling, which is
$\cE_g^x(2002)= 2,004,888.5938241700$, which failed the 10$^{th}$ and the 20$^{th}$
monotonicity tests, with $\veps_g^x(2012)-\veps_g^x(2002) = -0.0125431412$ and
$\veps_g^x(2022)-\veps_g^x(2002) = -0.012518560$.
	This time we obtain two upper bounds from \Ref{UPPERboundEgN} (for $n=10$ and $n=20$), and
the better one is for $n=10$, viz. 
\begin{equation}
 \cE_g(2002) 
\leq 
\textstyle{\frac{2002\cdot 2001}{2012\cdot 2011}}\, \cE_g^x(2012) = 1,954,640.745
\label{UPPERboundEg2002}
\end{equation}
which is also a considerable improvement over the listed $\cE_g^x(2002)$.
	Not surprisingly, in this long list of computer-experimental data our harvest is richer than in
the previous log list. 
	An additional nine valus of $\cE_g^x(N)$ failed one or more of the monotonicity tests
(indicated by $(n=\times)$ behind $\cE_g^x(N)$), namely 

$\cE_g^x(2531)=\phantom{1}3,204,550.3074368200\, (n=1;19)$

$\cE_g^x(2561)=\phantom{1}3,207,772.6856023400\, (n=1)$

$\cE_g^x(3362)=\phantom{1}5,543,845.2403717400\, (n=2)$
 
$\cE_g^x(3480)=\phantom{1}5,941,792.7610333500\, (n=1;2)$
 
$\cE_g^x(3663)=\phantom{1}6,586,476.2826423300\, (n=9)$ 

$\cE_g^x(3720)=\phantom{1}6,793,857.9289983900\, (n=1)$ 

$\cE_g^x(4000)=\phantom{1}7,860,293.8236758000\, (n=2)$ 

$\cE_g^x(4260)=\phantom{1}8,920,193.5261720900\, (n=2)$ 

$\cE_g^x(4620)=          10,498,739.0438109000\, (n=2;4)$.

\noindent
	We leave it to the interested reader to use \Ref{UPPERboundEgN} to compute upper bounds on $\cE_g(N)$ for
the $N$ showing in this table and the pertinent $\cE_g^x(N+n)$ at \cite{BCM}. 

	All the data in \cite{AWRTSDW,PerezGetal} pass the monotonicity test, but these data 
are few and far between so that it is actually quite unlikely for any of them to fail a monotonicity test.
	In fact,  \cite{AWRTSDW} in their table list only five different $N\in\{2,...,2500\}$(!), 
with two different $\cE_g^x(N)$ each; \cite{PerezGetal} point out that they found a lower energy 
$\cE_g^x(2472)$ than did \cite{AWRTSDW}.
	Also \cite{BCM} provided a putative value for $N=2472$, and
their $\cE_g^x(2472)=2,987,485.953(...)$ is even lower than the one in \cite{PerezGetal}, which is
$\cE_g^x(2472)=2,987,486.132$.
	Incidentally, there is also a moral here, to be told in the last section.

	To illustrate the monotonicity of the sequence 
$N\mapsto\veps_g(N)$ I plot the monotonically increasing sequence $N\mapsto\veps_g^x(N)\geq \veps_g(N)$, 
this time using the data in \cite{Sloanetal}. 
	For $N\leq 15$ also these two sequences are identical \cite{AtiyahSutcliffe}.
\bigskip

\epsfxsize=13.5cm
\epsfysize=5cm
\centerline{\epsffile{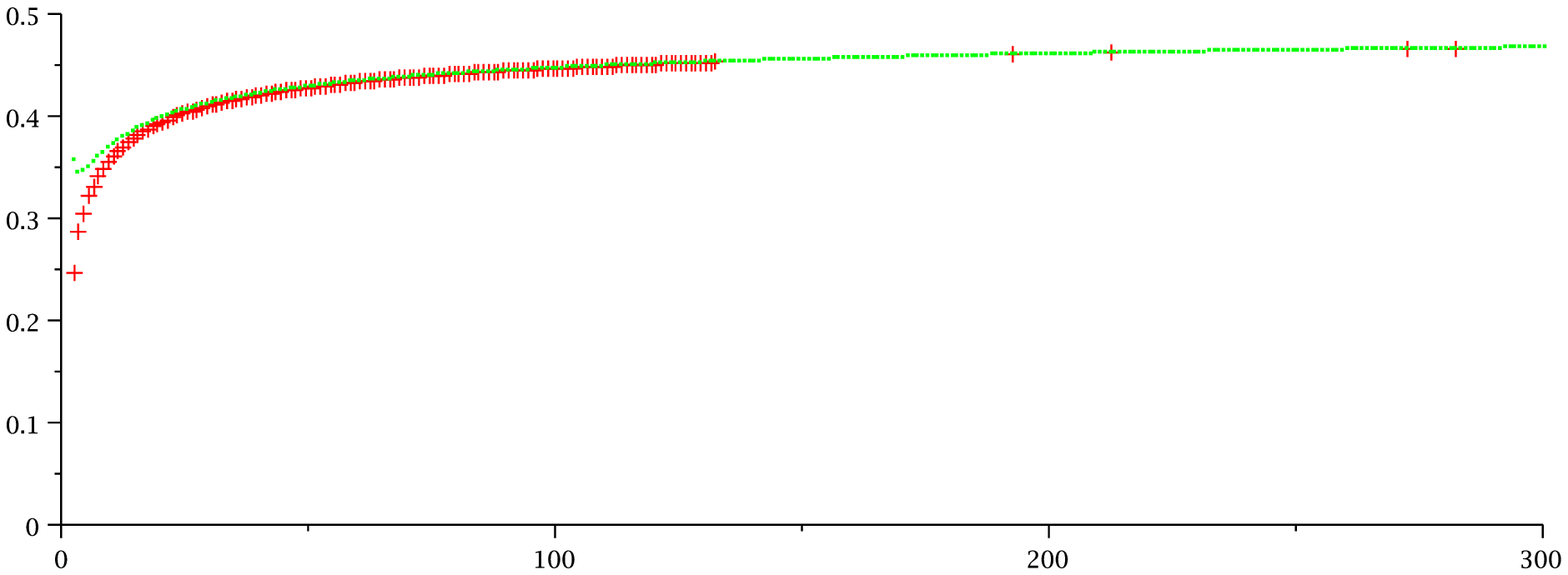}}
\begin{center}
{\scriptsize{\vskip-10pt
Computer-experimental pair-specific ground state energies $\veps_g^x(N)$ for $N$ point charges on \\
$\Sset^2$ with $r^{-1}\!$ Coulomb interactions, using the data of \cite{Sloanetal} (crosses). 
Also shown are the\\
\vskip-.2truecm
leading three terms in the conjectured asymptotic large $N$ approximation to $\veps_g(N)$ (dots).}}
\end{center}
\vskip-.2truecm
	Also shown in this diagram is the \emph{partially conjectured} large $N$ asymptotics 
$\veps_g(N)\asymp 
	\textstyle{\frac{1}{2}} - 0.55305(...) N^{-1/2} +\textstyle{\frac{1}{2}}N^{-1}+O(N^{-3/2})$, 
obtained through dividing by $N(N-1)$ the following large-$N$ asymptotic formula for $N\mapsto\cE_g(N)$,
\begin{equation}
 \cE_g(N) = a N^2 + b N^{3/2} + c N + dN^{1/2} + e + O(N^{-1/2}),
\label{asympCONJinvrS2}
\end{equation}
\vskip-.1truecm
\noindent
where $a=1/2$ is the only rigorously proven coefficient \cite{RSZa}, \cite{KuijlaarsSaff}, \cite{KieSpoCMP}, while 
it is conjectured \cite{RSZa} that  $c=0=e$ and
\begin{equation}
 b = 
3\textstyle\left(\frac{\sqrt{3}}{8\pi}\right)^{\scriptstyle{1/2}}\zeta\left(\frac{1}{2}\right)\sum_{k=0}^\infty 
\left( \frac{1}{\sqrt{3k+1}}-\frac{1}{\sqrt{3k+2}}\right) = - 0.55305...,
\end{equation}
\vskip-.1truecm
\noindent
whereas $d$ is estimated numerically in \cite{RSZb}.

\section{Many point charges on a 2-torus}
\noindent	
	For the perhaps most prominent non-spherical topology, the 2-torus $\Tset^2\subset\Rset^3$, 
we found data lists at \cite{BCM}.
	However, these lists are clearly preliminary.

\subsection{Two-dimensional Coulomb interactions}
\noindent
	Curiously, at \cite{BCM} putative ground state energies $\cE_g^x(N)$ are only listed for four 
different values of $N$, and all of them are obtained with different aspect ratios of the tori. 
	Hence, no meaningful monotonicity test can be applied.

	A visualization of a putative ground state configuration of $N$ point charges 
with logarithmic Coulomb interactions on a 2-torus can be found in \cite{HardinSaffONE} and on the
cover of that issue of the Notices, and also at \cite{Womersley}.

\subsection{Three-dimensional Coulomb interactions}
\noindent
	At \cite{BCM} one finds about 50 data of putative ground state energies $\cE_g^x(N)$
for the aspect ratio $1.414$, which are computed for sparcely placed $N$ up to $N=5000$.
	For sparsely placed data one would expect it to be less likely to find some which fail a 
monotonicity test. 
	However, the data $\cE_g^x(N)$ for nine $N$ failed a monotonicity test for one or more $n$, namely:

$\cE_g^x(15)\phantom{12}=\phantom{1,234,5}81.390479174\, (n=5;6;9;11;12)$

$\cE_g^x(30)\phantom{12}=\phantom{1,2345}331.088832684\, (n=2;5;6;7;10)$ 
 
$\cE_g^x(113)\phantom{1}=\phantom{1,23}5,370.892624565\, (n=4;7;38;49;78)$
 
$\cE_g^x(262)\phantom{1}=\phantom{1,2}28,287.128667479\,(n=38;238;290)$

$\cE_g^x(360)\phantom{1}=\phantom{1,2}53,857.158956562\,(n=140;192;200;694)$ 

$\cE_g^x(396)\phantom{1}=\phantom{1,2}66,660.796433247\, (n=104;156;164;598)$ 

$\cE_g^x(1000)=\phantom{12}418,396.928796506\, (n=363;4000)$

$\cE_g^x(1363)=\phantom{12}707,154.008010865\, (n=3637)$

$\cE_g^x(3500)= 5,174,438.587013800\,(n=1500) $.

\noindent
	Again, we leave it to the interested reader to use \Ref{UPPERboundEgN} and the above table 
to compute upper bounds on $\cE_g(N)$ from the pertinent data list at \cite{BCM}.
\section{Some variations on the theme}
\noindent
	Two- and three-dimensional Coulomb pair interactions and 2-sphere and 2-torus domains are  
merely the most prominent examples of pair interactions ${U}_\Lambda(\qV_i,\qV_j)$ and ${\rm d}$-dimensional 
domains $\Lambda$ to which Proposition \ref{prop:UperNsqMinusN} applies.
	Other, though physically less important, examples of interactions are ${\rm D}$-dimensional Coulomb 
interactions with ${\rm D}>3$, and more generally the so-called Riesz interactions, computed with  $U^{(s)}_\Lambda(\qV_i,\qV_j) = -\sign(s)|\qV_i-\qV_j|^{s}$, for any real $s<2$; see, e.g.
\cite{KuijlaarsSaff}, \cite{HardinSaffONE}, \cite{HardinSaffTWO}, \cite{KuijlaarsSaffSun}, \cite{BCEG}.
	The logarithmic Coulomb interaction is usually considered to be the special case $s=0$, in the
sense that $\lim_{s\downarrow 0} s^{-1}(|\qV_i-\qV_j|^{-s}-1)= -\ln |\qV_i-\qV_j|$;
	For $s=1$ the Riesz energy gives $U_\Lambda(\qV_i,\qV_j) = -|\qV_i-\qV_j|$, in which case
Proposition \ref{prop:UperNsqMinusN} may shed some new light on the question of the maximum average pair-wise 
distance of points in $\Lambda$ and related problems; beside the cited general survey articles, also see 
\cite{Beck}.
	Proposition \ref{prop:UperNsqMinusN} applies also to other bounded domains, in particular curves!
	I should emphasize that the logarithmic interactions between charges constrained to 
(planar) curves can be studied in quite some detail with complex variable techniques, see \cite{KBCDKNV}.
	Proposition \ref{prop:UperNsqMinusN} can easily be generalized to unbounded domains, with lower
semi-continuity replaced by another appropriate condition guaranteeing minimizing configurations for all $N$.
	A physically important example is $\Lambda=\Rset^3$ with 
$U_{\Rset^3}(\qV_i,\qV_j) =|\qV_i-\qV_j|^{-12}-|\qV_i-\qV_j|^{-6}$, 
which has minimizing $N$-body configurations for each $N$, known as Lennard-Jones clusters, 
see \cite{AtiyahSutcliffe} for a recent survey.
	At the expense of replacing the minimum by an infimum, pair interactions which are merely bounded below can 
be handled also, but minimizing sequences which don't converge to a minimizing configuration are perhaps less 
interesting.
\section{Summary}
\noindent
	The main purpose of this article is to draw attention to the monotonicity tests implied by Proposition
\ref{prop:UperNsqMinusN} and to emphasize the ups and downs of this monotonicity test family, not to report 
on an exhaustive series of such tests covering all available data. 
	In fact, having demonstrated the utility of these tests, it is much more efficient when they are being 
directly implemented in the computer experiments rather than being run by a third person afterwards.

	I still owe the reader the moral announced earlier.
	Originally I had analyzed the data of computer-experimental ground state energies 
$\cE_g^x(N)$ reported in \cite{AWRTSDW} which, divided by $N(N-1)$,
arranged themselves monotonically increasing when plotted vs. $N$; in fact, this prompted me to
conjecture and then prove Proposition \ref{prop:UperNsqMinusN}. 	
	But, as we saw, those data are not the correct ground state energies.
	With hindsight I was quite lucky, for such widely spaced  data almost inevitably form a 
pair-specific monotonic sequence.
	Had I hit upon a more closely spaced list of non-optimal data which would not have been pair-specifically
monotonic, I may not have conjectured the monotonicity in the first place!

\bigskip

\noindent
{\textbf{Acknowledgment}.} 
	This paper was written with support from the NSF under grant DMS-0807705.
	Any opinions expressed in this paper are entirely those of the author and not those of the NSF. 
	I like to thank my colleague A. Shadi Tahvildar-Zadeh for going through the 
painstaking process of verifying that those data which failed a monotonicity test
were really there and not made up by me.\footnote{Just in case 
		you  take a look and don't find those data anymore which failed a monotonicity test --- 
		there were some that did!
		Incidentally, I myself plan to try to improve on some data which failed a monotonicity
		test after registering on the interactive website~\cite{BCM}.}
	I also thank J\'ozsef Beck and Doron Zeilberger for their comments on Proposition 1, 
and Xiaoping Peng for helping me with the data processing.


\end{document}